\documentclass[reprint,english,preprintnumbers,amsmath,amssymb,aps,prl]{revtex4-2}
\usepackage{graphicx}
\usepackage{dsfont}
\usepackage{color}
\usepackage{multirow}
\usepackage{breakurl}

\usepackage{charter} 
\usepackage[charter]{mathdesign}

\definecolor{refcol}{RGB}{178,34,34}
\usepackage{microtype}
\usepackage[colorlinks,linkcolor=refcol,citecolor=refcol,urlcolor=refcol]{hyperref}

\graphicspath{{./figures/}}

\def\det{{\mathrm{det}}}

\def\Eq#1{Eq.~(\ref{#1})}

\def\Fig#1{Fig.~\ref{#1}}

\definecolor{blue}{rgb}{0,0,1}

\definecolor{green}{rgb}{0,1,0}

\definecolor{red}{rgb}{1,0,0}

\begin{document}

\title{Signatures of Moat Regimes in Heavy-Ion Collisions}

\author{Robert D.\ Pisarski}
\email[E-mail: ]{rob.pisarski@gmail.com}
\affiliation{Physics Department, Brookhaven National Laboratory,  Upton, NY 11973, USA} 

\author{Fabian Rennecke}
\email[E-mail: ]{frennecke@bnl.gov}
\affiliation{Condensed Matter Physics and Materials Science Division, Brookhaven National Laboratory, Upton, NY 11973, USA}

\begin{abstract}
  Heavy-ion collisions at small beam energies have the potential to reveal
  the rich phase structure of QCD at low temperature and nonzero density.
  In this case spatially modulated regimes with a "moat" spectrum can arise, 
  where the minimum of the energy is over a sphere at nonzero momentum.
  We show that if the matter created in heavy-ion collisions freezes out in such a regime,
  particle numbers and their correlations peak at nonzero, instead of zero, momentum.
  This effect is much more dramatic for multi-particle correlations than for single particle spectra.
  Our results can serve as a first guideline for a systematic search of spatially modulated phases in heavy-ion collisions.
\end{abstract}

\maketitle


{\bf Introduction --}\label{sec:intro}
The collisions of heavy ions at high energy, such
as at the Relativistic Heavy Ion Collider at BNL and the Large Hadron Collider at CERN, have
demonstrated experimentally the production of a qualitatively new state of matter.
This is possibly the transition to a quark-gluon plasma at a pseudocritical temperature $T \!\approx\! 154$~MeV
and net baryon chemical potential $\mu_B \!\approx\! 0$.  As the energy of the collision decreases, one probes
the phase diagram at lower $T$
and larger $\mu_B$ \cite{Fukushima:2010bq, Bzdak:2019pkr}.

Model studies suggest that regimes with periodic spatial modulations can occur high $\mu_B$. There, particles can have a "moat" spectrum, where the minimum of the energy
is not at zero momentum, but lies over a sphere at {\it non}zero spatial momentum. If the energy at the bottom of the moat is zero, an instability towards the formation of spatially inhomogeneous condensates might occur \cite{Buballa:2014tba}.
Studies in $1\!+\!1$ dimensional models that
are exactly soluble as the number of flavor $N_f \!\rightarrow\! \infty$ have shown that inhomogeneous chiral condensates {\it always} arise at low temperature and large density
\cite{Basar:2008im,Basar:2008ki,Basar:2009fg}.
How the moat spectrum manifests itself depends upon the internal symmetry.
For these soluble models the global symmetry is either discrete ($Z(2)$) or $O(2)$,
and the spatially inhomogeneous condensate derived at $N_f = \infty$ should persist at
finite $N_f$
\footnote{Speaking strictly the results derived only apply at infinite $N_f$. Due to the infrared fluctuations in two spacetime dimensions, at finite $N_f$ the $O(2)$ symmetry cannot spontaneously break; similarly, for a discrete symmetry at nonzero temperature. This is special to two spacetime dimensions, as in more than two spacetime dimensions, the spatially inhomogeneous condensate indicated by mean field theory does persist for either a discrete or $O(2)$ symmetry. There are numerous examples in condensed matter, such as smectic-C liquid crystals, etc., which confirm this.}.

Recent analysis, however, suggests that this is radically changed by the presence of Goldstone bosons.
Consider an $O(N)$ model, so that when $N > 2$, the spatially inhomogeneous condensate (or "chiral spiral")
has $N-2$ Goldstone modes.
About the (simplest) chiral spiral, unexpectedly the Goldstone bosons have {\it zero} energy at the bottom
of the moat. Since this occurs at nonzero momentum, it is a double pole,
and generates severe linear infrared divergences 
\footnote{We emphasize that this is different from the Landau-Peierls theorem, where fluctuations in transverse spatial direction around a one-dimensional inhomogeneous condensate lead to logarithmic divergences \cite{Landau:1980mil}.}.  
A solution at infinite $N$ demonstrates that a dynamical mass for the Goldstone bosons is generated at the
bottom of the moat, and disorders the would-be chiral spirals
into a type of quantum pion liquid (Q$\pi$L)
\cite{Pisarski:2020gkx,Pisarski:2020dnx,Pisarski:2021aoz,Tsvelik:2021ccp}.
This is a disordered phase, where unlike a chiral spiral, there is {\it no} preferred direction.
Because the minimum of the energy is at nonzero momentum, it is still associated with periodic spatial modulations.
For QCD, where two light flavors are described
by an $O(4)$ model, this suggests that instead of a condensate of chiral spirals, there is a Q$\pi$L 
\footnote{In the chiral limit, for three flavors the Goldstone bosons should disorder the chiral spiral into a quantum liquid of pions and kaons. What happens in QCD, when the strange quark is much heavier than the up and down quarks, is not obvious. The kaon fluctuations could either form a quantum liquid, like the pions, or a real kink crystal; see Ref. \cite{Pisarski:2020dnx}.}.

In this Letter we consider the experimental signatures of a moat spectrum.
The difference between a normal homogeneous phase, where the smallest energy is at zero momentum,
and for a moat regime, where it is at nonzero momentum, should be direct.
Certainly the particle distribution with a moat spectrum should be
peaked at nonzero momentum \cite{Pisarski:2020gkx}. 
We argue that this elementary feature can leave distinctive signatures in the production of particles
and their correlations, which are measurable in heavy-ion collisions.

Both in QCD \cite{Fu:2019hdw} and 
in models \cite{Basar:2008im,Basar:2008ki,Basar:2009fg,Buballa:2014tba}
a moat spectrum can arise in a large region of the phase diagram.
Thus, the matter created in heavy-ion collisions at low beam energies
is likely to traverse a moat regime if it exists at large $\mu_B$.
For simplicity, we assume that freeze-out occurs in this region.
This does not exclude the possibility that the signatures discussed here could also occur
if the system is in a moat regime at earlier stages of its evolution.
In any case, if particles at sufficiently low momenta can be measured, we predict a peak
in the momentum dependence of particle numbers and correlations at \emph{nonzero} momentum.
This is a clear signature of spatial modulations in dense QCD,
and could signal either a Q$\pi$L or a precursor to the formation of an inhomogeneous phase.

\vspace{5pt}
{\bf Moat Spectrum --}\label{sec:piQpiL}
We focus on the subsector of the QCD action at low energy which contains quark bilinears, and is described by
an $O(N)$ field $\vec{\phi}$; for $N=4$, this contains the $\sigma$ meson and three pions for two light flavors:
\begin{align}\label{eq:lag}
  \mathcal{L}= \frac{1}{2} \big(\!\partial_0 \vec{\phi}\big)^2 - \frac{Z}{2} \big(\!\partial_i \vec{\phi}\big)^2
  - \frac{W}{2} \big(\!\partial_i^2 \vec{\phi}\big)^2 - \frac{m_{\rm eff}^2}{2}\, \vec{\phi}^2\, + \ldots
\end{align}
There are higher order terms in the Lagrangian, such as a quartic coupling
$\sim (\vec{\phi}^2)^2$, but all that matters for our analysis are the terms
quadratic in $\vec{\phi}$. In vacuum, by Lorentz invariance $Z=1$, but a medium is not Lorentz invariant,
and so $Z$ can be less than unity.
In a moat regime, one has the more dramatic feature that $Z$ is negative; then terms of quartic order
in the spatial momentum are
required for stability at large momentum.  
As an effective theory at low momenta, we therefore add such a term, whose coefficient, $W$, must be positive.

\begin{figure}[t]
\centering
\includegraphics[width=.9\columnwidth]{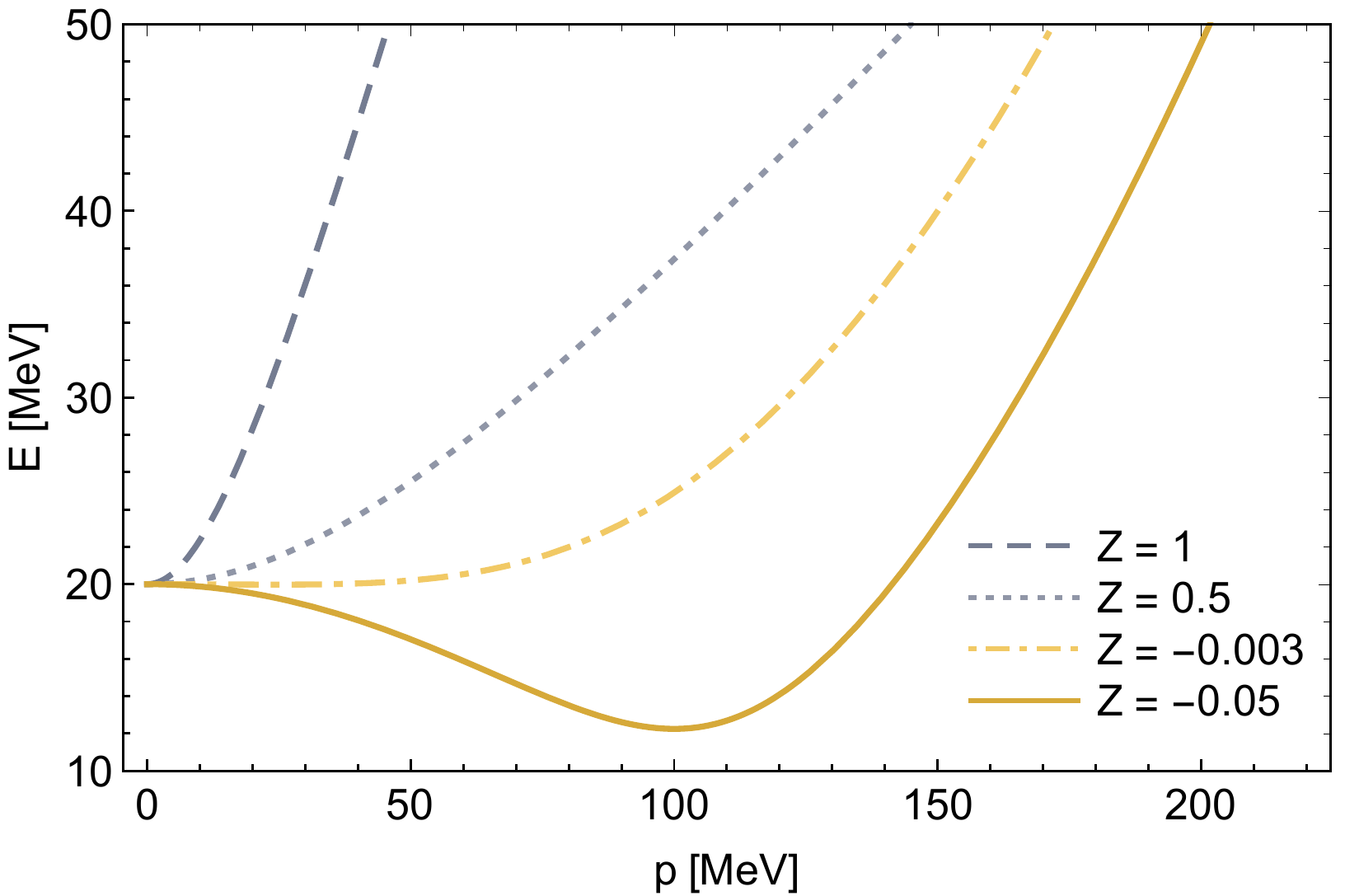}
\caption{Dispersion relations in the normal ($Z>0$, gray lines) and moat ($Z<0$, yellow lines) regime as functions of the spatial momentum. As opposed to the normal phase, the energy is minimal at a finite momentum in the moat regime. We have chosen $W = 2.5\, \rm{GeV}^{-2}$ when $Z<0$.}
\label{fig:disp}
\end{figure}

If $Z\!<\!0$, then classically an instability towards the formation of an inhomogeneous condensate, e.g., a chiral spiral $\vec{\phi} = \sigma_0 \big( \cos(p_{\rm min}\, z),\,\sin(p_{\rm min}\, z), 0, \dots,0 \big)$, can occur, where $p_{\rm min}^2$ is the bottom of the moat. However, for $N\rightarrow\infty$ it can be shown rigorously that a double pole in the Goldstone modes at $p_{\rm min}$ always disorders the inhomogeneous condensate \cite{Pisarski:2020dnx, Pisarski:2021aoz, Tsvelik:2021ccp}.
Instead of such condensate, the system is in a Q$\pi$L phase, which is characterized by spatial correlations of the field $\phi$
which fall-off with the typical exponential times an oscillatory function.
While this is certain at large-$N$, the Q$\pi$L should arise whenever there are Goldstone bosons for a chiral spiral,
$N > 2$ \cite{Pisarski:2020dnx}, and certainly includes $N=4$.

In the moat regime, $Z<0$, and the energy
\begin{align}\label{eq:Ephi}
E_\phi(\mathbf{p}^2) = \sqrt{Z\mathbf{p}^2 + W (\mathbf{p}^2)^2 + m_{\rm eff}^2} \, ,
\end{align}
has a minimum at nonzero momentum, $p_{\rm min}^2 = -Z/(2W)$.
$p_{\rm min}$ is related to the wavenumber of the underlying spatial modulation.
Examples for different $Z$ are shown in \Fig{fig:disp}.
We note that for sufficiently large momenta, the energy should reduce to the one in vacuum.
Admittedly, the values which we choose in \Fig{fig:disp} and henceforth are illustrative
and do not follow from a detailed model.
However, the results of \cite{Fu:2019hdw} indicate that deeper in the moat regime $Z$ could even take much larger
negative values than the ones explored here
\footnote{In Ref.\ \cite{Fu:2019hdw} indications of a moat regime were found for $\mu_B \gtrsim 420\, {\rm MeV}$ and $T \gtrsim T_c(\mu_B)$. In this case, $Z$ and $W$ in \Eq{eq:Ephi} parametrize the momentum dependence of quark-antiquark scattering in meson (specifically pion) exchange channels. Hence, moat regimes in the deconfined phase correspond to enhanced scattering at $p_{\rm min}$ in these channels. If this feature persists in the hadronic phase, it directly translates into a meson dispersion as in \Eq{eq:Ephi}. However, this has not been analyzed in \cite{Fu:2019hdw}.}.
More importantly, it is natural to expect that $p_{\rm min}$ is related to a low-energy scale $\lesssim \Lambda_{\rm QCD}$.
We further require $E_\phi(\mathbf{p}^2) > 0$ for all $\mathbf{p}^2$ in order to avoid an instability.
In any case, our analysis builds on the intuitive fact that when $p_{\rm min}$ is nonzero,
the production of particles with nonzero momentum is favored.

\vspace{5pt}
{\bf Particle Spectra --}\label{sec:spec}
We consider the production of particles in heavy-ion collisions in more detail.
Matter produced in heavy ion collisions expands rapidly
and cools until the system is so dilute that particles freeze out:
elastic and inelastic scattering cease and hadrons stream to the detector.
Hence, we need to describe particles on the freeze-out surface,
where, if the system passes through a moat regime during its expansion,
particles can carry signatures of this regime.
Here, we restrict ourselves to the case where the freeze-out occurs in the moat regime.
Studying the case where the system traverses a moat regime during earlier stages of its evolution,
but freezes out in a different phase, is beyond the scope of the present work.

The number of particles on the freeze-out surface can be computed by using
the formula of Cooper and Frye \cite{Cooper:1974mv}.  This assumes 
free (quasi)-particles with an ordinary dispersion relation, $(p^0)^2 = \mathbf{p}^2 + m^2$.
Here we need to generalize this result to describe particles with an arbitrary dispersion relation
\cite{PisarskiRennecke};
see Refs.\ \cite{Grossi:2020ezz, Erschfeld:2020blf, Grossi:2021gqi} for related discussions.
In thermal equilibrium, the Wigner function,
$F_\phi(p)$, of the field $\phi$ is proportional to the spectral function $\rho_\phi$,
\begin{align}\label{eq:wigner}
F_\phi(p) = 2\pi\, \rho_\phi(p^0,\mathbf{p})\, n_B(p_0)\, ,
\end{align}
where $n_B(p_0)$ is the Bose-Einstein statistical distribution function.
With this, the momentum dependent spectrum of $\phi$ on the freeze-out surface for particle $i$ is given by
\begin{align}\label{eq:fospec}
n_i \equiv \frac{d^3 N_\phi}{d \mathbf{p}_i^3} 
=\frac{2}{(2\pi)^3} \int_\Sigma\, d\Sigma_\mu \int\! \frac{d p^0_i}{2\pi}\,p_i^\mu\, \Theta(\breve p^0_i)\, F_\phi(\breve p_i)\,.
\end{align}
$\Sigma$ is the freeze-out surface, which is given by a three-dimensional submanifold of four-dimensional spacetime, determined by the hypersurface
with a fixed temperature, the freeze-out temperature $T_f$. The coordinates on this surface are $w^{1,2,3}$, and the induced metric
$G_{ij} = g^{\mu\nu} (dx^\mu/dw^i)(dx^\nu/dw^j)$, where $g^{\mu\nu} = \rm{diag}(1,-1,-1,-1)$ is the Minkowski metric.
The integral measure over the hypersurface is therefore given by $d\Sigma^\mu = \hat v^\mu\, \sqrt{|\det G|}\, d^3w$,
with the normal vector $\hat v^\mu = v^\mu/|v|$ given by the generalized
cross product of the tangent vectors on $\Sigma$,
$v^\mu = \bar\epsilon^{\mu\alpha\beta\gamma}\, (dx_\alpha/dw^1)(dx_\beta/dw^2)(dx_\gamma/dw^3)$, where
$\bar\epsilon^{\mu\alpha\beta\gamma}$ is the covariant Levi-Civita tensor.

The frequency $\breve{p}_0$  and the spatial momentum $\breve{\mathbf{p}}$ are boosted by the fluid four-velocity $u^\mu$ of the medium at freeze-out,
$\breve p_0 = u^\mu p_\mu$ and $\breve{\mathbf p}^2\, = \Delta^{\mu\nu} p_\mu p_\nu$,
where $\Delta^{\mu\nu} = u^\mu u^\nu - g^{\mu\nu}$ is the projection operator for spatial directions transverse to $u^\mu$.
Note that  $F_\phi(\breve p) = 2\pi\, \rho_\phi(\breve p_0 , \breve{\mathbf p}^2)\, n_B(\breve p_0)$, so that the spectral function and the
thermal distribution are evaluated at the boosted momenta.
The Heaviside function $\Theta(\breve p_0)$ ensures that only particles with positive energy contribute.
We implicitly assumed rotational invariance of the spectral function, since it is not broken in the moat regime, although
\Eq{eq:fospec} is valid regardless.

\Eq{eq:fospec} is obtained by integrating the covariant particle number current over the freeze-out surface. If
the spectral function of the particle is known, for a given parametrization of the freeze-out surface and the fluid velocity,
particle spectra can readily be computed. The freeze-out surface and the fluid velocity are obtained from the hydrodynamic
evolution of the system.
For simplicity we parametrize ideal hydrodynamics in terms of a boost-invariant blast wave model \cite{Schnedermann:1993ws, Florkowski:2010zz},
assuming that the velocity of the radial expansion of the system grows proportional
to the radial distance, and that freeze-out happens at a fixed proper time, $\tau_f$.
The freeze-out curve can then be described in terms of the longitudinal rapidity $\eta_\parallel = \rm{artanh}(z/t)$,
the radius $r = \sqrt{x^2+y^2}$,
and the angle $\phi = \arctan(y/x)$, where $z$ is the direction of the beam. We use the parametrization of Ref.\ \cite{Teaney:2003kp}, which yields for the fluid velocity
\begin{align}
u^\mu = \big( u^\tau \cosh \eta_\parallel,\, u^r \cos\phi,\, u^r \sin\phi,\, u^\tau \sinh \eta_\parallel \big)\, ,
\end{align}
with the radial flow velocity $u^r = \bar u\, \frac{r}{\bar R}\, \Theta(\bar R - r)$ and $u^\tau = \sqrt{1+(u^r)^2}$.
The fluid velocity has two free parameters, the radial velocity, $\bar u$, and the fireball radius, $\bar R$.
The vector normal to the freeze-out surface
is $\hat v^\mu = (\cosh\eta_\parallel, 0, 0, \sinh\eta_\parallel )$
and the determinant of the induced metric is $\sqrt{|\det G|} = \tau_f r\, \Theta(\bar R -r)$. 

The relevant physics of the moat regime is encoded in the spectral function $\rho_\phi(p^0,{\mathbf p})$.
While we use a boost-invariant blast wave model for the fluid-dynamical part of our
calculation, particles with a moat spectrum clearly break boost invariance.
We compute the quasi-particle spectral function from the free Lagrangian in \Eq{eq:lag}, 
\begin{align}\label{eq:rhophi}
\rho_\phi(\breve p_0 , \breve{\mathbf p}^2) = \rm{sign}(\breve p_0)\, \delta\big[(\breve p_0)^2-E_\phi^2(\breve{\mathbf p}^2)\big] \, ,
\end{align}
where $E_\phi$ is the energy given in \Eq{eq:Ephi}.
Note that since the energy is quartic in the spatial momenta, performing the
frequency integration in \Eq{eq:fospec} even with this simple spectral function is nontrivial.
The momenta in \Eq{eq:rhophi} are boosted, so the on-shell condition $(\breve p_0)^2 = E_\phi^2(\breve{\mathbf p}^2) $
is in general a quartic equation in the rest frame momentum, $p_0$.
Of the four possible complex-valued solutions, only the ones which give real, positive $\breve{p}_0$
contribute to \Eq{eq:fospec}.
For ordinary free bosons, i.e., with $Z=1$ and $W=0$ in \Eq{eq:lag}, this computation is trivial and
\Eq{eq:fospec} reduces to the well-known Cooper-Frye formula. 

The spectral function in \Eq{eq:rhophi} implies that the smallest energy $\breve{p}_0$
that contributes to the spectrum is given by the minimum of $E_\phi^2(\breve{\mathbf p}^2)$.
In the moat regime this minimum is at a nonzero spatial momentum.
Since the thermal distribution function gives the largest contribution at the lowest energy in \Eq{eq:fospec},
most particles are produced at nonzero momentum.

As shown in Fig. (21) of Ref.\ \cite{Fu:2019hdw},
there are large regions in the $T-\mu_B$ plane in which 
the wave function renormalization $Z$ is negative, indicating a moat regime such as the Q$\pi$L.
The parameters $Z$, $W$ and $m_{\rm eff}$, and thus the spectral function, are all functions of
$T$ and $\mu_B$.  These can be computed from first principles by methods such as
the functional renormalization group following Refs.\ \cite{Fu:2019hdw}
and, e.g., \cite{Tripolt:2013jra, Jung:2016yxl, Tripolt:2020irx}.
For illustrative purposes, we choose parameters
where the main features of the particle spectrum are apparent. In the homogeneous phase
we choose $W = 0$ and $Z>0$, while in the moat regime we take $W = 2.5\, \rm{GeV}^{-2}$ and  $Z<0$.
In both cases we take $m_{\rm eff} = 20$\,MeV.
We stress again that these are illustrative, yet reasonable parameters, as explained above.

To fix the thermodynamic and blast-wave parameters,
we first pick a beam-energy of $\sqrt{s} = 5$\,GeV, since it yields a system which is dense enough to traverse
the preliminary moat region in Ref.\ \cite{Fu:2019hdw}.
The corresponding freeze-out temperature and chemical potential from statistical model fits in Ref.\ \cite{Andronic:2017pug}
are $T_f = 115$\,MeV and $\mu_{B,f} = 536$\,MeV.
From blast-wave fits to experimental data at different beam energies in Ref.\ \cite{Zhang:2016tbf},
we read-off the blast-wave parameters $\bar u = 0.3$, $\bar R = 8$\,fm and $\tau_f = 5$\,fm/c.

\begin{figure}[t]
\centering
\includegraphics[width=.9\columnwidth]{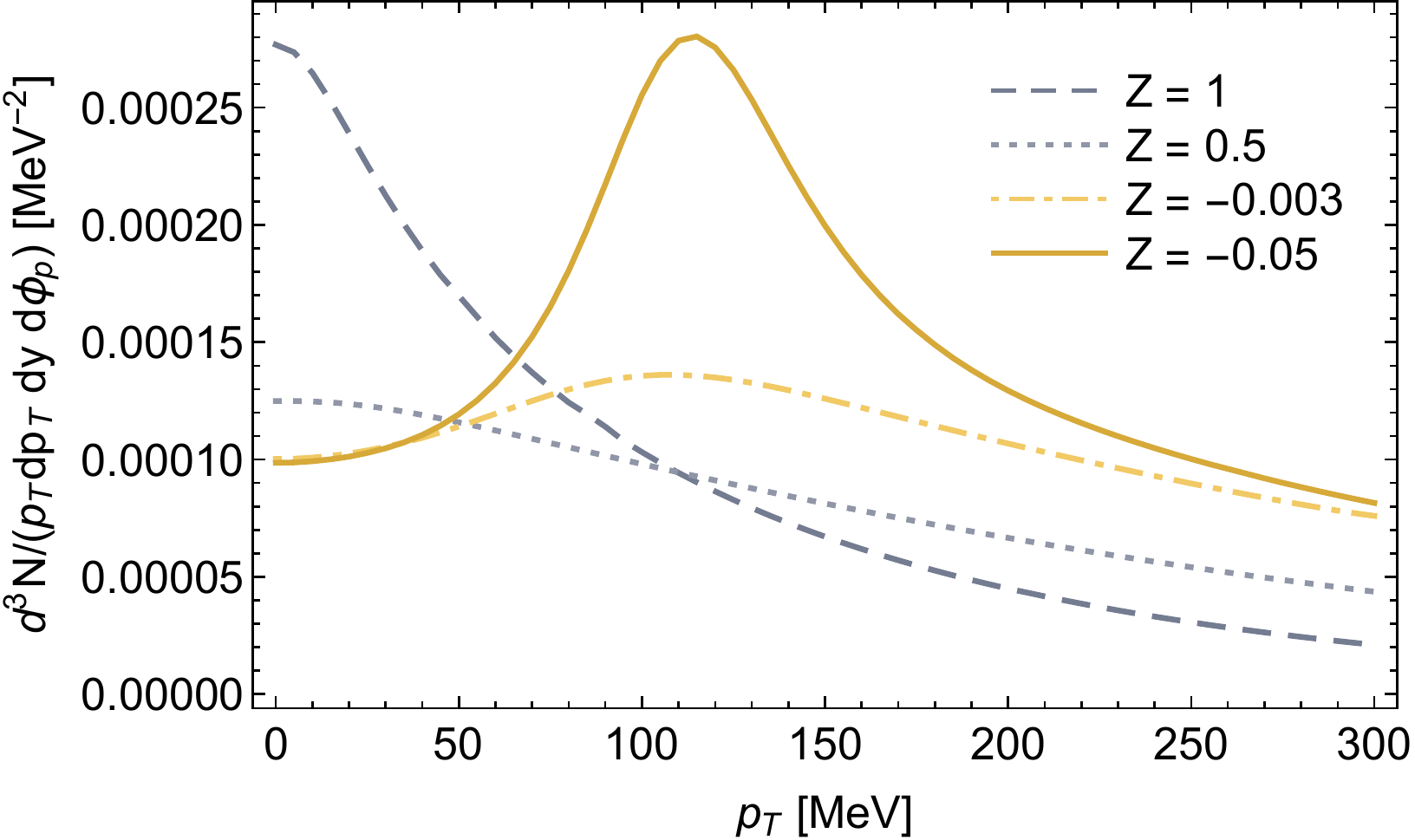}
\caption{Comparison between the transverse momentum spectra in the normal ($Z>0$, gray lines)
  and moat ($Z<0$, yellow lines) regime.
  While most particles are produced at zero momentum and the spectrum decays
  exponentially with $p_T$ in the normal phase, it has a distinct maximum at nonzero $p_T$ in the moat regime.}
\label{fig:pTspec}
\end{figure}

Having fixed all unknowns in \Eq{eq:fospec}, we show the resulting particle spectra for different $Z$
in \Fig{fig:pTspec} as a function of the transverse momentum $p_T = \sqrt{p_x^2+p_y^2}$.
By rotational invariance we can set the azimuthal momentum angle $\phi_p = \arctan(p_y/p_x) =0$,
and take the momentum space rapidity $y = {\rm arctanh}(p_z/p_0) =0$.
The numerical results show the behavior anticipated above.
Since as a function of momentum the energy is shallower for smaller $|Z|$, so are the spectra.
Further, in the moat regime the minimal energy,
$E_{\rm min} = \sqrt{m^2_{\rm eff} - Z^2/4 W}$, is less than the minimal energy in the normal phase, $m_{\rm eff}$.
Thus, as $Z$ becomes more negative, the energy is lowered and particle production increased at nonzero momentum.
Since the radial boost velocity increases towards the edges of the fireball, the $p_T$ spectrum shows a
slightly washed-out peak at finite momentum corresponding to a boosted $p_{\rm min}$.
We expect that the moat dispersion eventually turns into the normal one at large momenta.
Since this is not captured in our model, the large $p_T$ behavior should not be taken too seriously here.

In summary, a monotonously increasing $p_T$ spectrum at low momenta, and a peak at finite momentum,
are clear signatures of a moat regime.

\vspace{5pt}
{\bf Particle Number Correlations --}\label{sec:corr}
In addition to the particle spectra,
a moat spectrum should also affect the momentum dependence of correlations in particle number.
It is reasonable to anticipate that correlations are even more sensitive to moat regimes than the spectra.
These can be computed just by considering single-particle correlations, since while there is
no long range order in the system,
the particles are modified by their propagation in the medium.  This is in contrast to a critical endpoint,
where long-range correlations, and hence genuine multi-particle effects, appear.

Particle number correlations are given by a straightforward generalization of \Eq{eq:fospec},
\begin{align}\label{eq:corgen}
\nonumber
\Bigg\langle \prod_{i} \frac{d^3 N_\phi}{d \mathbf{p}_i^3}  \Bigg\rangle &= \Bigg[\prod_i \frac{2}{(2\pi)^3} \int\! d\Sigma^\mu_i \int\! \frac{dp_i^0}{2\pi}\,(p_i)_\mu\, \Theta(\breve p^0_i)\Bigg]\,\\[2pt] 
&\quad\times\bigg\langle\prod_i  F_\phi(\breve p_i) \bigg\rangle\,.
\end{align}
Nontrivial correlations arise when $F_\phi$ fluctuates. Here, we focus on thermodynamic fluctuations
in the spacetime-dependent temperature, baryon chemical potential and fluid
velocity of the expanding medium. These are summarized by the quantity
\begin{align}
\kappa_i^\mu(x) = \big( T(x),\,\mu_B(x),\, u^\mu(x) \big)_i\,.
\end{align}
$\kappa$ fluctuates around its mean value,
$\langle \kappa \rangle = \bar\kappa$, i.e., $\kappa = \bar\kappa + \delta\kappa$,
which implies $\langle \delta\kappa \rangle = 0$.
Fluctuations in $\kappa$ induce those in $F_\phi$.
Assuming that the fluctuations are small, $F_\phi$ can be expanded in powers of $\delta\kappa$,
and correlations of Wigner functions translate directly into correlations of thermodynamic parameters.
For example, expanding up to second order in $\delta\kappa$, the \emph{connected} two-point correlation of $F_\phi$ is
\begin{align}
\big\langle F_\phi\,F_\phi  \big\rangle_c 
= \frac{\partial F_\phi}{\partial\kappa_i^\mu}  \frac{\partial F_\phi}{\partial\kappa_j^\nu}\bigg|_{\bar\kappa}
\big\langle \delta\kappa_i^\mu \delta\kappa_j^\nu \big\rangle +\mathcal{O}\big(\delta\kappa^3\big)\,,
\end{align}
where we used that the connected two-point function is determined only by the fluctuations,
$\langle F_\phi^2 \rangle_c = \langle F_\phi^2 \rangle - \langle F_\phi \rangle^2$.

To compute the thermodynamic average, we generalize the classic treatment of thermodynamic fluctuations
in Ref.\ \cite{Landau:1980mil}, using unpublished results of Ref.\
\footnote{Private communication with S.\ Floerchinger (2016), and unpublished work presented
by S.\ Floerchinger at the workshop
{"\emph{Functional Methods in Strongly Correlated Systems}"} in Hirschegg, Austria, on April 4, 2019}.
We construct the generating functional for thermodynamic correlations on the freeze-out surface,
\begin{align}\label{eq:genfunc}
  {\rm e}^{W[J]} = \int\!\!\!\mathcal{D}\kappa \exp\! \int\!\!\!
  d\Sigma_\mu  \left[\Delta s^\mu(w) + \ J(w)_{i\nu}\, \hat{v}^\mu\delta\kappa^\nu_i(w)  \right] \; .
\end{align}
Taking $n$ derivatives of $W[J]$ with respect to the source $J$ and then setting $J=0$ yields the connected
$n$-point function $\sim \langle (\delta\kappa)^n \rangle_c$.
$\Delta s^\mu$ is the change in the entropy current density due to fluctuations
in each fluid cell on the freeze-out surface,
i.e., the difference between the entropy current density of the subsystem away from equilibrium specified by
$\bar\kappa + \delta\kappa$ and that in equilibrium at $\bar\kappa$.
For small fluctuations, it is given by the second-order correction of the
expansion of the entropy current in terms of the thermodynamic fields,
\begin{equation}
  \Delta s^\mu = \frac{1}{2} \frac{\partial^2 s^\mu}{\partial \kappa_i^\nu \partial \kappa_j^\rho}\bigg|_{\bar\kappa}\,
  \delta\kappa_i^\nu \delta\kappa_j^\rho \; .
\end{equation}
%
\begin{figure}[t]
\centering
\includegraphics[width=.9\columnwidth]{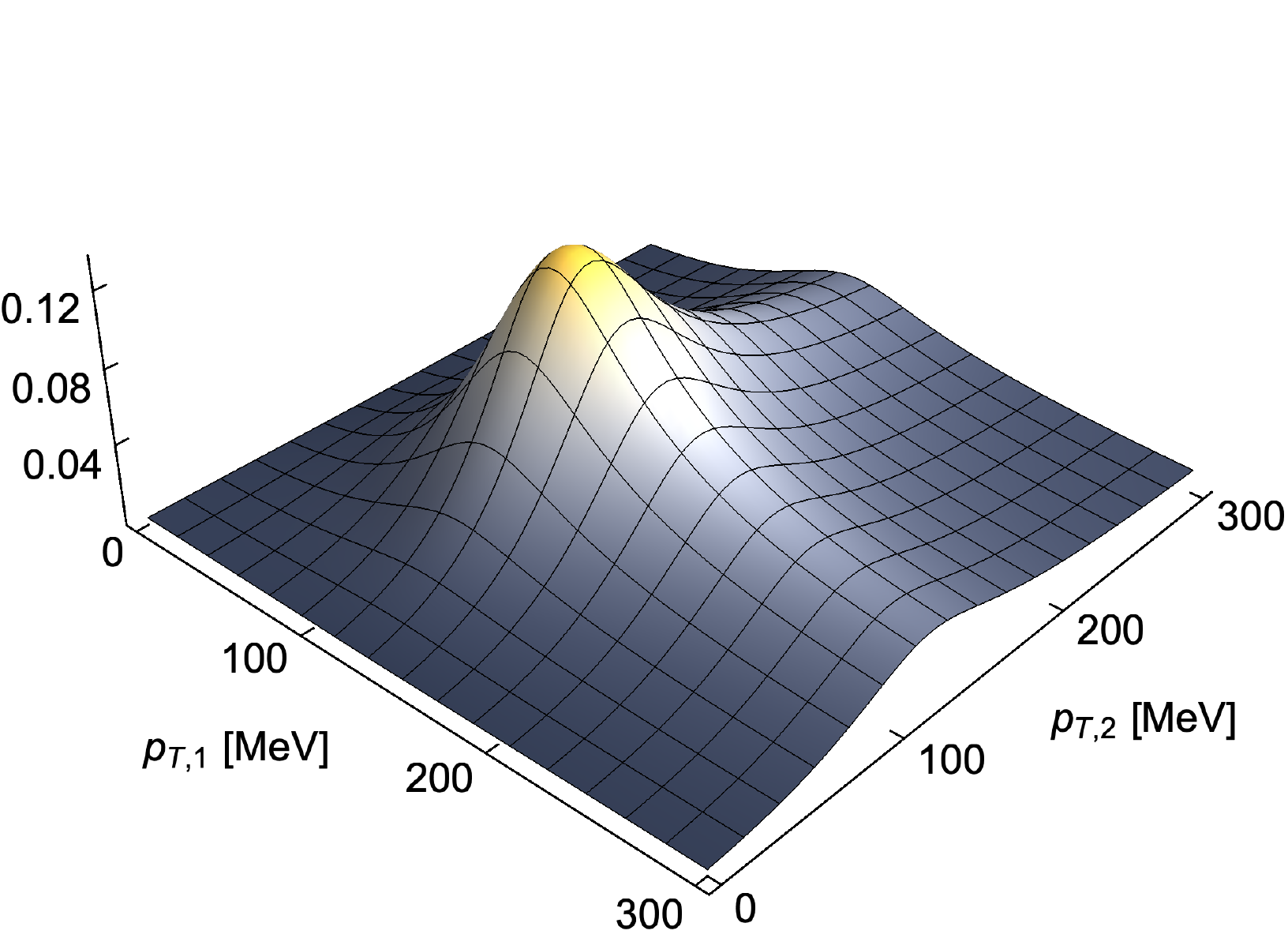}
\caption{In the moat regime, the connected two-particle correlation function normalized by the mean spectra,
$n_{12}/(n_1 n_2)$, Eqs. (\ref{eq:fospec}) and (\ref{eq:conncorr}), as a function of $p_T$.
The peak and the two ridges show that fluctuations are enhanced at a momentum related to the wavenumber of
the spatial modulation of the moat regime. The correlation in the normal phase is
basically flat and much smaller compared to the peak in the moat regime, by at least $\sim 10^{-2}$.}
\label{fig:conncorr}
\end{figure}
Using the covariant form of the Euler relation, $ds^\mu = (u_\nu/T)\, dT^{\mu\nu} - (\mu_B/T)\ dn^\mu$,
where $T^{\mu\nu} = \epsilon\, u^\mu u^\nu + p \Delta^{\mu\nu}$ is the ideal energy-momentum tensor, we find
\begin{align}
\hat v_\mu \Delta s^\mu = -\frac{1}{2} \delta\kappa_{i\mu}(w)\, \mathcal{F}^{\mu\nu}_{ij}(w)\, \delta\kappa_{j\nu}(w)\,.
\end{align}
The thermodynamic fluctuations are encoded in the fluctuation matrix
\begin{align}\label{eq:Fdef}
\mathcal{F}^{\mu\nu}_{ij} &= \frac{1}{T}\!
\begin{pmatrix}
\hat u\, \dfrac{\partial s}{\partial T}& \hat u\, \dfrac{\partial s}{\partial\mu_B} & s \hat v^\nu\\[10pt]
\hat u\, \dfrac{\partial s}{\partial\mu_B} & \hat u\, \dfrac{\partial n}{\partial \mu_B} & n \hat v^\nu\\[10pt]
 s \hat v^\mu &  n \hat v^\mu & -\hat u\,(T s + \mu_B n) g^{\mu\nu}
\end{pmatrix}_{ij}\,,
\end{align}
with $\hat u = \hat v_\mu u^\mu$.
As $\mathcal{F}(w)$ is local, so are the thermodynamic fluctuations.   Assuming that the fluctuations are small,
only Gaussian fluctuations contribute, and the only nontrivial correlator in \Eq{eq:genfunc} 
is the two-point function:
\begin{align}\label{eq:conncorr}
\nonumber
n_{12} &\equiv \Bigg\langle \frac{d^3 N_\phi}{d\mathbf{p}_1^3} \frac{d^3 N_\phi}{d\mathbf{p}_2^3}  \Bigg\rangle_c = \\[2pt]
\nonumber
&\quad \frac{4}{(2\pi)^6} \int\! d\Sigma^\mu \int\!
\frac{dp_1^0}{2\pi} \frac{dp_2^0}{2\pi}\,(p_1)_\mu\,
(\hat v \!\cdot\! p_2)\,\Theta(\breve p^0_1)\, \Theta(\breve p^0_2)\\[2pt]
&\quad\times \Bigg(\frac{\partial F_\phi(\breve p_1)}{\partial\kappa^\rho_{i}} 
\frac{\partial F_\phi(\breve p_2)}{\partial\kappa^\sigma_{j}}\Bigg)\Bigg|_{\bar\kappa}\,
\left( \mathcal{F}_{ij}^{\rho\sigma}(w)\right)^{-1}\,.
\end{align}
All other correlation functions can be obtained via Wick's theorem, as products of two-point functions.
As before, we can evaluate this in the blast wave approximation with the Wigner function
of \Eq{eq:wigner} and the spectral function of \Eq{eq:rhophi}.
To specify the thermodynamic quantities on the freeze-out surface,
we use a hadron resonance gas model
\cite{BraunMunzinger:2003zd} which includes known particles and resonances with masses up to 2.6\,GeV \cite{Zyla:2020zbs}.

In \Fig{fig:conncorr} we show the resulting connected two-point correlator as a function of
transverse momenta in the moat regime, again with illustrative parameters
$Z = -0.05$, $W = 2.5\,{\rm GeV}^{-2}$ and $m_{\rm eff} = 20$\,MeV.
As before, we set $\phi_p$ and $y$ to zero and note that little changes qualitatively at nonzero rapidity.
There is a clear enhancement of the fluctuations at a nonzero momentum, in contrast to the normal phase,
where {\it no} such structure occurs.
Comparing with the single particle spectrum in \Fig{fig:pTspec}, measuring the fluctuations through
the two-particle correlation function yields a {\it much} greater enhancement in the moat regime, as
compared to the normal phase.  While our model is a gross simplification of the hadronic cocktail which
contributes to a realistic simulation, any ordinary spectra will uniformly peak at zero momentum, and yield
a tiny and flat two particle distribution, as compared to \Fig{fig:conncorr}.

\vspace{5pt}
{\bf Conclusions --}\label{sec:conc}
We have shown how a moat regime leaves its imprint in particle spectra and their correlations.
The key signature is that both show a characteristic peak at a
nonzero momentum at the bottom of the moat.
Thus, $p_T$-resolved measurements of these quantities in heavy-ion collisions at low energy are an
efficient way of detecting moat regimes.
While the formalism which we developed is general,
we used simple models for the evolution of matter produced in heavy-ion
collisions. Still, our results capture the essential physical features and therefore provide a
first guide for how and where to look for novel phases in the QCD phase diagram.
Experimentally, the main problem is to measure the particle spectrum at sufficiently
low momentum.  In the Beam Energy Scan II at RHIC, momenta down to $\sim 60\, {\rm MeV}$ will
be measured.  Hopefully the novel signatures which we propose may appear.

A more detailed analysis will be presented separately \cite{PisarskiRennecke}.

\vspace{5pt}
{\it Acknowledgments --}
We thank H. Caines, W. Christie, S. Floerchinger, E. Grossi, and L. Ruan for discussions and comments.
This research was supported by the U.S. Department of Energy 
under contract DE-SC0012704 and by B.N.L. under the Lab Directed Research and Development program 18-036.

\bibliography{moat}

\begin{thebibliography}{32}%
\makeatletter
\providecommand \@ifxundefined [1]{%
 \@ifx{#1\undefined}
}%
\providecommand \@ifnum [1]{%
 \ifnum #1\expandafter \@firstoftwo
 \else \expandafter \@secondoftwo
 \fi
}%
\providecommand \@ifx [1]{%
 \ifx #1\expandafter \@firstoftwo
 \else \expandafter \@secondoftwo
 \fi
}%
\providecommand \natexlab [1]{#1}%
\providecommand \enquote  [1]{``#1''}%
\providecommand \bibnamefont  [1]{#1}%
\providecommand \bibfnamefont [1]{#1}%
\providecommand \citenamefont [1]{#1}%
\providecommand \href@noop [0]{\@secondoftwo}%
\providecommand \href [0]{\begingroup \@sanitize@url \@href}%
\providecommand \@href[1]{\@@startlink{#1}\@@href}%
\providecommand \@@href[1]{\endgroup#1\@@endlink}%
\providecommand \@sanitize@url [0]{\catcode `\\12\catcode `\$12\catcode
  `\&12\catcode `\#12\catcode `\^12\catcode `\_12\catcode `\%12\relax}%
\providecommand \@@startlink[1]{}%
\providecommand \@@endlink[0]{}%
\providecommand \url  [0]{\begingroup\@sanitize@url \@url }%
\providecommand \@url [1]{\endgroup\@href {#1}{\urlprefix }}%
\providecommand \urlprefix  [0]{URL }%
\providecommand \Eprint [0]{\href }%
\providecommand \doibase [0]{https://doi.org/}%
\providecommand \selectlanguage [0]{\@gobble}%
\providecommand \bibinfo  [0]{\@secondoftwo}%
\providecommand \bibfield  [0]{\@secondoftwo}%
\providecommand \translation [1]{[#1]}%
\providecommand \BibitemOpen [0]{}%
\providecommand \bibitemStop [0]{}%
\providecommand \bibitemNoStop [0]{.\EOS\space}%
\providecommand \EOS [0]{\spacefactor3000\relax}%
\providecommand \BibitemShut  [1]{\csname bibitem#1\endcsname}%
\let\auto@bib@innerbib\@empty
\bibitem [{\citenamefont {Fukushima}\ and\ \citenamefont
  {Hatsuda}(2011)}]{Fukushima:2010bq}%
  \BibitemOpen
  \bibfield  {author} {\bibinfo {author} {\bibfnamefont {K.}~\bibnamefont
  {Fukushima}}\ and\ \bibinfo {author} {\bibfnamefont {T.}~\bibnamefont
  {Hatsuda}},\ }\bibfield  {title} {\bibinfo {title} {{The phase diagram of
  dense QCD}},\ }\href {https://doi.org/10.1088/0034-4885/74/1/014001}
  {\bibfield  {journal} {\bibinfo  {journal} {Rept. Prog. Phys.}\ }\textbf
  {\bibinfo {volume} {74}},\ \bibinfo {pages} {014001} (\bibinfo {year}
  {2011})},\ \Eprint {https://arxiv.org/abs/1005.4814} {arXiv:1005.4814
  [hep-ph]} \BibitemShut {NoStop}%
\bibitem [{\citenamefont {Bzdak}\ \emph {et~al.}(2020)\citenamefont {Bzdak},
  \citenamefont {Esumi}, \citenamefont {Koch}, \citenamefont {Liao},
  \citenamefont {Stephanov},\ and\ \citenamefont {Xu}}]{Bzdak:2019pkr}%
  \BibitemOpen
  \bibfield  {author} {\bibinfo {author} {\bibfnamefont {A.}~\bibnamefont
  {Bzdak}}, \bibinfo {author} {\bibfnamefont {S.}~\bibnamefont {Esumi}},
  \bibinfo {author} {\bibfnamefont {V.}~\bibnamefont {Koch}}, \bibinfo {author}
  {\bibfnamefont {J.}~\bibnamefont {Liao}}, \bibinfo {author} {\bibfnamefont
  {M.}~\bibnamefont {Stephanov}},\ and\ \bibinfo {author} {\bibfnamefont
  {N.}~\bibnamefont {Xu}},\ }\bibfield  {title} {\bibinfo {title} {{Mapping the
  Phases of Quantum Chromodynamics with Beam Energy Scan}},\ }\href
  {https://doi.org/10.1016/j.physrep.2020.01.005} {\bibfield  {journal}
  {\bibinfo  {journal} {Phys. Rept.}\ }\textbf {\bibinfo {volume} {853}},\
  \bibinfo {pages} {1} (\bibinfo {year} {2020})},\ \Eprint
  {https://arxiv.org/abs/1906.00936} {arXiv:1906.00936 [nucl-th]} \BibitemShut
  {NoStop}%
\bibitem [{\citenamefont {Buballa}\ and\ \citenamefont
  {Carignano}(2015)}]{Buballa:2014tba}%
  \BibitemOpen
  \bibfield  {author} {\bibinfo {author} {\bibfnamefont {M.}~\bibnamefont
  {Buballa}}\ and\ \bibinfo {author} {\bibfnamefont {S.}~\bibnamefont
  {Carignano}},\ }\bibfield  {title} {\bibinfo {title} {{Inhomogeneous chiral
  condensates}},\ }\href {https://doi.org/10.1016/j.ppnp.2014.11.001}
  {\bibfield  {journal} {\bibinfo  {journal} {Prog. Part. Nucl. Phys.}\
  }\textbf {\bibinfo {volume} {81}},\ \bibinfo {pages} {39} (\bibinfo {year}
  {2015})},\ \Eprint {https://arxiv.org/abs/1406.1367} {arXiv:1406.1367
  [hep-ph]} \BibitemShut {NoStop}%
\bibitem [{\citenamefont {Basar}\ and\ \citenamefont
  {Dunne}(2008{\natexlab{a}})}]{Basar:2008im}%
  \BibitemOpen
  \bibfield  {author} {\bibinfo {author} {\bibfnamefont {G.}~\bibnamefont
  {Basar}}\ and\ \bibinfo {author} {\bibfnamefont {G.~V.}\ \bibnamefont
  {Dunne}},\ }\bibfield  {title} {\bibinfo {title} {{Self-consistent
  crystalline condensate in chiral Gross-Neveu and Bogoliubov-de Gennes
  systems}},\ }\href {https://doi.org/10.1103/PhysRevLett.100.200404}
  {\bibfield  {journal} {\bibinfo  {journal} {Phys. Rev. Lett.}\ }\textbf
  {\bibinfo {volume} {100}},\ \bibinfo {pages} {200404} (\bibinfo {year}
  {2008}{\natexlab{a}})},\ \Eprint {https://arxiv.org/abs/0803.1501}
  {arXiv:0803.1501 [hep-th]} \BibitemShut {NoStop}%
\bibitem [{\citenamefont {Basar}\ and\ \citenamefont
  {Dunne}(2008{\natexlab{b}})}]{Basar:2008ki}%
  \BibitemOpen
  \bibfield  {author} {\bibinfo {author} {\bibfnamefont {G.}~\bibnamefont
  {Basar}}\ and\ \bibinfo {author} {\bibfnamefont {G.~V.}\ \bibnamefont
  {Dunne}},\ }\bibfield  {title} {\bibinfo {title} {{A Twisted Kink Crystal in
  the Chiral Gross-Neveu model}},\ }\href
  {https://doi.org/10.1103/PhysRevD.78.065022} {\bibfield  {journal} {\bibinfo
  {journal} {Phys. Rev.}\ }\textbf {\bibinfo {volume} {D78}},\ \bibinfo {pages}
  {065022} (\bibinfo {year} {2008}{\natexlab{b}})},\ \Eprint
  {https://arxiv.org/abs/0806.2659} {arXiv:0806.2659 [hep-th]} \BibitemShut
  {NoStop}%
\bibitem [{\citenamefont {Basar}\ \emph {et~al.}(2009)\citenamefont {Basar},
  \citenamefont {Dunne},\ and\ \citenamefont {Thies}}]{Basar:2009fg}%
  \BibitemOpen
  \bibfield  {author} {\bibinfo {author} {\bibfnamefont {G.}~\bibnamefont
  {Basar}}, \bibinfo {author} {\bibfnamefont {G.~V.}\ \bibnamefont {Dunne}},\
  and\ \bibinfo {author} {\bibfnamefont {M.}~\bibnamefont {Thies}},\ }\bibfield
   {title} {\bibinfo {title} {{Inhomogeneous Condensates in the Thermodynamics
  of the Chiral NJL(2) model}},\ }\href
  {https://doi.org/10.1103/PhysRevD.79.105012} {\bibfield  {journal} {\bibinfo
  {journal} {Phys. Rev.}\ }\textbf {\bibinfo {volume} {D79}},\ \bibinfo {pages}
  {105012} (\bibinfo {year} {2009})},\ \Eprint
  {https://arxiv.org/abs/0903.1868} {arXiv:0903.1868 [hep-th]} \BibitemShut
  {NoStop}%
\bibitem [{Note1()}]{Note1}%
  \BibitemOpen
  \bibinfo {note} {Speaking strictly the results derived only apply at infinite
  $N_f$. Due to the infrared fluctuations in two spacetime dimensions, at
  finite $N_f$ the $O(2)$ symmetry cannot spontaneously break; similarly, for a
  discrete symmetry at nonzero temperature. This is special to two spacetime
  dimensions, as in more than two spacetime dimensions, the spatially
  inhomogeneous condensate indicated by mean field theory does persist for
  either a discrete or $O(2)$ symmetry. There are numerous examples in
  condensed matter, such as smectic-C liquid crystals, etc., which confirm
  this.}\BibitemShut {Stop}%
\bibitem [{Note2()}]{Note2}%
  \BibitemOpen
  \bibinfo {note} {We emphasize that this is different from the Landau-Peierls
  theorem, where fluctuations in transverse spatial direction around a
  one-dimensional inhomogeneous condensate lead to logarithmic divergences
  \cite {Landau:1980mil}.}\BibitemShut {Stop}%
\bibitem [{\citenamefont {Pisarski}\ \emph {et~al.}(2021)\citenamefont
  {Pisarski}, \citenamefont {Rennecke}, \citenamefont {Tsvelik},\ and\
  \citenamefont {Valgushev}}]{Pisarski:2020gkx}%
  \BibitemOpen
  \bibfield  {author} {\bibinfo {author} {\bibfnamefont {R.~D.}\ \bibnamefont
  {Pisarski}}, \bibinfo {author} {\bibfnamefont {F.}~\bibnamefont {Rennecke}},
  \bibinfo {author} {\bibfnamefont {A.}~\bibnamefont {Tsvelik}},\ and\ \bibinfo
  {author} {\bibfnamefont {S.}~\bibnamefont {Valgushev}},\ }\bibfield  {title}
  {\bibinfo {title} {{The Lifshitz Regime and its Experimental Signals}},\
  }\href {https://doi.org/10.1016/j.nuclphysa.2020.121910} {\bibfield
  {journal} {\bibinfo  {journal} {Nucl. Phys. A}\ }\textbf {\bibinfo {volume}
  {1005}},\ \bibinfo {pages} {121910} (\bibinfo {year} {2021})},\ \Eprint
  {https://arxiv.org/abs/2005.00045} {arXiv:2005.00045 [nucl-th]} \BibitemShut
  {NoStop}%
\bibitem [{\citenamefont {Pisarski}\ \emph {et~al.}(2020)\citenamefont
  {Pisarski}, \citenamefont {Tsvelik},\ and\ \citenamefont
  {Valgushev}}]{Pisarski:2020dnx}%
  \BibitemOpen
  \bibfield  {author} {\bibinfo {author} {\bibfnamefont {R.~D.}\ \bibnamefont
  {Pisarski}}, \bibinfo {author} {\bibfnamefont {A.~M.}\ \bibnamefont
  {Tsvelik}},\ and\ \bibinfo {author} {\bibfnamefont {S.}~\bibnamefont
  {Valgushev}},\ }\bibfield  {title} {\bibinfo {title} {{How transverse thermal
  fluctuations disorder a condensate of chiral spirals into a quantum spin
  liquid}},\ }\href {https://doi.org/10.1103/PhysRevD.102.016015} {\bibfield
  {journal} {\bibinfo  {journal} {Phys. Rev. D}\ }\textbf {\bibinfo {volume}
  {102}},\ \bibinfo {pages} {016015} (\bibinfo {year} {2020})},\ \Eprint
  {https://arxiv.org/abs/2005.10259} {arXiv:2005.10259 [hep-ph]} \BibitemShut
  {NoStop}%
\bibitem [{\citenamefont {Pisarski}(2021)}]{Pisarski:2021aoz}%
  \BibitemOpen
  \bibfield  {author} {\bibinfo {author} {\bibfnamefont {R.~D.}\ \bibnamefont
  {Pisarski}},\ }\bibfield  {title} {\bibinfo {title} {{Remarks on nuclear
  matter: how an $\omega_0$ condensate can spike the speed of sound, and a
  model of $Z(3)$ baryons}},\ }\href@noop {} {\  (\bibinfo {year} {2021})},\
  \Eprint {https://arxiv.org/abs/2101.05813} {arXiv:2101.05813 [nucl-th]}
  \BibitemShut {NoStop}%
\bibitem [{\citenamefont {Tsvelik}\ and\ \citenamefont
  {Pisarski}(2021)}]{Tsvelik:2021ccp}%
  \BibitemOpen
  \bibfield  {author} {\bibinfo {author} {\bibfnamefont {A.~M.}\ \bibnamefont
  {Tsvelik}}\ and\ \bibinfo {author} {\bibfnamefont {R.~D.}\ \bibnamefont
  {Pisarski}},\ }\bibfield  {title} {\bibinfo {title} {{Low energy physics of
  interacting bosons with a moat spectrum, and the implications for condensed
  matter and cold nuclear matter}},\ }\href@noop {} {\  (\bibinfo {year}
  {2021})},\ \Eprint {https://arxiv.org/abs/2103.15835} {arXiv:2103.15835
  [nucl-th]} \BibitemShut {NoStop}%
\bibitem [{Note3()}]{Note3}%
  \BibitemOpen
  \bibinfo {note} {In the chiral limit, for three flavors the Goldstone bosons
  should disorder the chiral spiral into a quantum liquid of pions and kaons.
  What happens in QCD, when the strange quark is much heavier than the up and
  down quarks, is not obvious. The kaon fluctuations could either form a
  quantum liquid, like the pions, or a real kink crystal; see Ref. \cite
  {Pisarski:2020dnx}.}\BibitemShut {Stop}%
\bibitem [{\citenamefont {Fu}\ \emph {et~al.}(2020)\citenamefont {Fu},
  \citenamefont {Pawlowski},\ and\ \citenamefont {Rennecke}}]{Fu:2019hdw}%
  \BibitemOpen
  \bibfield  {author} {\bibinfo {author} {\bibfnamefont {W.-j.}\ \bibnamefont
  {Fu}}, \bibinfo {author} {\bibfnamefont {J.~M.}\ \bibnamefont {Pawlowski}},\
  and\ \bibinfo {author} {\bibfnamefont {F.}~\bibnamefont {Rennecke}},\
  }\bibfield  {title} {\bibinfo {title} {{QCD phase structure at finite
  temperature and density}},\ }\href
  {https://doi.org/10.1103/PhysRevD.101.054032} {\bibfield  {journal} {\bibinfo
   {journal} {Phys. Rev. D}\ }\textbf {\bibinfo {volume} {101}},\ \bibinfo
  {pages} {054032} (\bibinfo {year} {2020})},\ \Eprint
  {https://arxiv.org/abs/1909.02991} {arXiv:1909.02991 [hep-ph]} \BibitemShut
  {NoStop}%
\bibitem [{Note4()}]{Note4}%
  \BibitemOpen
  \bibinfo {note} {In Ref.\ \cite {Fu:2019hdw} indications of a moat regime
  were found for $\mu _B \gtrsim 420\protect \tmspace +\thinmuskip {.1667em}
  {\protect \rm MeV}$ and $T \gtrsim T_c(\mu _B)$. In this case, $Z$ and $W$ in
  Eq.~(\ref {eq:Ephi}) parametrize the momentum dependence of quark-antiquark
  scattering in meson (specifically pion) exchange channels. Hence, moat
  regimes in the deconfined phase correspond to enhanced scattering at
  $p_{\protect \rm min}$ in these channels. If this feature persists in the
  hadronic phase, it directly translates into a meson dispersion as in
  Eq.~(\ref {eq:Ephi}). However, this has not been analyzed in \cite
  {Fu:2019hdw}.}\BibitemShut {Stop}%
\bibitem [{\citenamefont {Cooper}\ and\ \citenamefont
  {Frye}(1974)}]{Cooper:1974mv}%
  \BibitemOpen
  \bibfield  {author} {\bibinfo {author} {\bibfnamefont {F.}~\bibnamefont
  {Cooper}}\ and\ \bibinfo {author} {\bibfnamefont {G.}~\bibnamefont {Frye}},\
  }\bibfield  {title} {\bibinfo {title} {{Comment on the Single Particle
  Distribution in the Hydrodynamic and Statistical Thermodynamic Models of
  Multiparticle Production}},\ }\href {https://doi.org/10.1103/PhysRevD.10.186}
  {\bibfield  {journal} {\bibinfo  {journal} {Phys. Rev. D}\ }\textbf {\bibinfo
  {volume} {10}},\ \bibinfo {pages} {186} (\bibinfo {year} {1974})}\BibitemShut
  {NoStop}%
\bibitem [{\citenamefont {Pisarski}\ and\ \citenamefont
  {Rennecke}(2021)}]{PisarskiRennecke}%
  \BibitemOpen
  \bibfield  {author} {\bibinfo {author} {\bibfnamefont {R.~D.}\ \bibnamefont
  {Pisarski}}\ and\ \bibinfo {author} {\bibfnamefont {F.}~\bibnamefont
  {Rennecke}},\ }\href@noop {} {\bibfield  {journal} {\bibinfo  {journal} {{in
  preparation}}\ } (\bibinfo {year} {2021})}\BibitemShut {NoStop}%
\bibitem [{\citenamefont {Grossi}\ \emph {et~al.}(2020)\citenamefont {Grossi},
  \citenamefont {Soloviev}, \citenamefont {Teaney},\ and\ \citenamefont
  {Yan}}]{Grossi:2020ezz}%
  \BibitemOpen
  \bibfield  {author} {\bibinfo {author} {\bibfnamefont {E.}~\bibnamefont
  {Grossi}}, \bibinfo {author} {\bibfnamefont {A.}~\bibnamefont {Soloviev}},
  \bibinfo {author} {\bibfnamefont {D.}~\bibnamefont {Teaney}},\ and\ \bibinfo
  {author} {\bibfnamefont {F.}~\bibnamefont {Yan}},\ }\bibfield  {title}
  {\bibinfo {title} {{Transport and hydrodynamics in the chiral limit}},\
  }\href {https://doi.org/10.1103/PhysRevD.102.014042} {\bibfield  {journal}
  {\bibinfo  {journal} {Phys. Rev. D}\ }\textbf {\bibinfo {volume} {102}},\
  \bibinfo {pages} {014042} (\bibinfo {year} {2020})},\ \Eprint
  {https://arxiv.org/abs/2005.02885} {arXiv:2005.02885 [hep-th]} \BibitemShut
  {NoStop}%
\bibitem [{\citenamefont {Erschfeld}\ \emph {et~al.}(2020)\citenamefont
  {Erschfeld}, \citenamefont {Floerchinger},\ and\ \citenamefont
  {Rupprecht}}]{Erschfeld:2020blf}%
  \BibitemOpen
  \bibfield  {author} {\bibinfo {author} {\bibfnamefont {A.}~\bibnamefont
  {Erschfeld}}, \bibinfo {author} {\bibfnamefont {S.}~\bibnamefont
  {Floerchinger}},\ and\ \bibinfo {author} {\bibfnamefont {M.}~\bibnamefont
  {Rupprecht}},\ }\bibfield  {title} {\bibinfo {title} {{General relativistic
  nonideal fluid equations for dark matter from a truncated cumulant
  expansion}},\ }\href {https://doi.org/10.1103/PhysRevD.102.063520} {\bibfield
   {journal} {\bibinfo  {journal} {Phys. Rev. D}\ }\textbf {\bibinfo {volume}
  {102}},\ \bibinfo {pages} {063520} (\bibinfo {year} {2020})},\ \Eprint
  {https://arxiv.org/abs/2005.12923} {arXiv:2005.12923 [hep-ph]} \BibitemShut
  {NoStop}%
\bibitem [{\citenamefont {Grossi}\ \emph {et~al.}(2021)\citenamefont {Grossi},
  \citenamefont {Soloviev}, \citenamefont {Teaney},\ and\ \citenamefont
  {Yan}}]{Grossi:2021gqi}%
  \BibitemOpen
  \bibfield  {author} {\bibinfo {author} {\bibfnamefont {E.}~\bibnamefont
  {Grossi}}, \bibinfo {author} {\bibfnamefont {A.}~\bibnamefont {Soloviev}},
  \bibinfo {author} {\bibfnamefont {D.}~\bibnamefont {Teaney}},\ and\ \bibinfo
  {author} {\bibfnamefont {F.}~\bibnamefont {Yan}},\ }\bibfield  {title}
  {\bibinfo {title} {{Soft pions and transport near the chiral critical
  point}},\ }\href@noop {} {\  (\bibinfo {year} {2021})},\ \Eprint
  {https://arxiv.org/abs/2101.10847} {arXiv:2101.10847 [nucl-th]} \BibitemShut
  {NoStop}%
\bibitem [{\citenamefont {Schnedermann}\ \emph {et~al.}(1993)\citenamefont
  {Schnedermann}, \citenamefont {Sollfrank},\ and\ \citenamefont
  {Heinz}}]{Schnedermann:1993ws}%
  \BibitemOpen
  \bibfield  {author} {\bibinfo {author} {\bibfnamefont {E.}~\bibnamefont
  {Schnedermann}}, \bibinfo {author} {\bibfnamefont {J.}~\bibnamefont
  {Sollfrank}},\ and\ \bibinfo {author} {\bibfnamefont {U.~W.}\ \bibnamefont
  {Heinz}},\ }\bibfield  {title} {\bibinfo {title} {{Thermal phenomenology of
  hadrons from 200-A/GeV S+S collisions}},\ }\href
  {https://doi.org/10.1103/PhysRevC.48.2462} {\bibfield  {journal} {\bibinfo
  {journal} {Phys. Rev. C}\ }\textbf {\bibinfo {volume} {48}},\ \bibinfo
  {pages} {2462} (\bibinfo {year} {1993})},\ \Eprint
  {https://arxiv.org/abs/nucl-th/9307020} {arXiv:nucl-th/9307020} \BibitemShut
  {NoStop}%
\bibitem [{\citenamefont {Florkowski}(2010)}]{Florkowski:2010zz}%
  \BibitemOpen
  \bibfield  {author} {\bibinfo {author} {\bibfnamefont {W.}~\bibnamefont
  {Florkowski}},\ }\href@noop {} {\emph {\bibinfo {title} {{Phenomenology of
  Ultra-Relativistic Heavy-Ion Collisions}}}}\ (\bibinfo  {publisher} {{World
  Scientific Publishing Company}},\ \bibinfo {year} {2010})\BibitemShut
  {NoStop}%
\bibitem [{\citenamefont {Teaney}(2003)}]{Teaney:2003kp}%
  \BibitemOpen
  \bibfield  {author} {\bibinfo {author} {\bibfnamefont {D.}~\bibnamefont
  {Teaney}},\ }\bibfield  {title} {\bibinfo {title} {{The Effects of viscosity
  on spectra, elliptic flow, and HBT radii}},\ }\href
  {https://doi.org/10.1103/PhysRevC.68.034913} {\bibfield  {journal} {\bibinfo
  {journal} {Phys. Rev. C}\ }\textbf {\bibinfo {volume} {68}},\ \bibinfo
  {pages} {034913} (\bibinfo {year} {2003})},\ \Eprint
  {https://arxiv.org/abs/nucl-th/0301099} {arXiv:nucl-th/0301099} \BibitemShut
  {NoStop}%
\bibitem [{\citenamefont {Tripolt}\ \emph {et~al.}(2014)\citenamefont
  {Tripolt}, \citenamefont {Strodthoff}, \citenamefont {von Smekal},\ and\
  \citenamefont {Wambach}}]{Tripolt:2013jra}%
  \BibitemOpen
  \bibfield  {author} {\bibinfo {author} {\bibfnamefont {R.-A.}\ \bibnamefont
  {Tripolt}}, \bibinfo {author} {\bibfnamefont {N.}~\bibnamefont {Strodthoff}},
  \bibinfo {author} {\bibfnamefont {L.}~\bibnamefont {von Smekal}},\ and\
  \bibinfo {author} {\bibfnamefont {J.}~\bibnamefont {Wambach}},\ }\bibfield
  {title} {\bibinfo {title} {{Spectral Functions for the Quark-Meson Model
  Phase Diagram from the Functional Renormalization Group}},\ }\href
  {https://doi.org/10.1103/PhysRevD.89.034010} {\bibfield  {journal} {\bibinfo
  {journal} {Phys. Rev. D}\ }\textbf {\bibinfo {volume} {89}},\ \bibinfo
  {pages} {034010} (\bibinfo {year} {2014})},\ \Eprint
  {https://arxiv.org/abs/1311.0630} {arXiv:1311.0630 [hep-ph]} \BibitemShut
  {NoStop}%
\bibitem [{\citenamefont {Jung}\ \emph {et~al.}(2017)\citenamefont {Jung},
  \citenamefont {Rennecke}, \citenamefont {Tripolt}, \citenamefont {von
  Smekal},\ and\ \citenamefont {Wambach}}]{Jung:2016yxl}%
  \BibitemOpen
  \bibfield  {author} {\bibinfo {author} {\bibfnamefont {C.}~\bibnamefont
  {Jung}}, \bibinfo {author} {\bibfnamefont {F.}~\bibnamefont {Rennecke}},
  \bibinfo {author} {\bibfnamefont {R.-A.}\ \bibnamefont {Tripolt}}, \bibinfo
  {author} {\bibfnamefont {L.}~\bibnamefont {von Smekal}},\ and\ \bibinfo
  {author} {\bibfnamefont {J.}~\bibnamefont {Wambach}},\ }\bibfield  {title}
  {\bibinfo {title} {{In-Medium Spectral Functions of Vector- and Axial-Vector
  Mesons from the Functional Renormalization Group}},\ }\href
  {https://doi.org/10.1103/PhysRevD.95.036020} {\bibfield  {journal} {\bibinfo
  {journal} {Phys. Rev. D}\ }\textbf {\bibinfo {volume} {95}},\ \bibinfo
  {pages} {036020} (\bibinfo {year} {2017})},\ \Eprint
  {https://arxiv.org/abs/1610.08754} {arXiv:1610.08754 [hep-ph]} \BibitemShut
  {NoStop}%
\bibitem [{\citenamefont {Tripolt}\ \emph {et~al.}(2020)\citenamefont
  {Tripolt}, \citenamefont {Rischke}, \citenamefont {von Smekal},\ and\
  \citenamefont {Wambach}}]{Tripolt:2020irx}%
  \BibitemOpen
  \bibfield  {author} {\bibinfo {author} {\bibfnamefont {R.-A.}\ \bibnamefont
  {Tripolt}}, \bibinfo {author} {\bibfnamefont {D.~H.}\ \bibnamefont
  {Rischke}}, \bibinfo {author} {\bibfnamefont {L.}~\bibnamefont {von
  Smekal}},\ and\ \bibinfo {author} {\bibfnamefont {J.}~\bibnamefont
  {Wambach}},\ }\bibfield  {title} {\bibinfo {title} {{Fermionic excitations at
  finite temperature and density}},\ }\href
  {https://doi.org/10.1103/PhysRevD.101.094010} {\bibfield  {journal} {\bibinfo
   {journal} {Phys. Rev. D}\ }\textbf {\bibinfo {volume} {101}},\ \bibinfo
  {pages} {094010} (\bibinfo {year} {2020})},\ \Eprint
  {https://arxiv.org/abs/2003.11871} {arXiv:2003.11871 [hep-ph]} \BibitemShut
  {NoStop}%
\bibitem [{\citenamefont {Andronic}\ \emph {et~al.}(2018)\citenamefont
  {Andronic}, \citenamefont {Braun-Munzinger}, \citenamefont {Redlich},\ and\
  \citenamefont {Stachel}}]{Andronic:2017pug}%
  \BibitemOpen
  \bibfield  {author} {\bibinfo {author} {\bibfnamefont {A.}~\bibnamefont
  {Andronic}}, \bibinfo {author} {\bibfnamefont {P.}~\bibnamefont
  {Braun-Munzinger}}, \bibinfo {author} {\bibfnamefont {K.}~\bibnamefont
  {Redlich}},\ and\ \bibinfo {author} {\bibfnamefont {J.}~\bibnamefont
  {Stachel}},\ }\bibfield  {title} {\bibinfo {title} {{Decoding the phase
  structure of QCD via particle production at high energy}},\ }\href
  {https://doi.org/10.1038/s41586-018-0491-6} {\bibfield  {journal} {\bibinfo
  {journal} {Nature}\ }\textbf {\bibinfo {volume} {561}},\ \bibinfo {pages}
  {321} (\bibinfo {year} {2018})},\ \Eprint {https://arxiv.org/abs/1710.09425}
  {arXiv:1710.09425 [nucl-th]} \BibitemShut {NoStop}%
\bibitem [{\citenamefont {Zhang}\ \emph {et~al.}(2016)\citenamefont {Zhang},
  \citenamefont {Ma}, \citenamefont {Chen},\ and\ \citenamefont
  {Zhong}}]{Zhang:2016tbf}%
  \BibitemOpen
  \bibfield  {author} {\bibinfo {author} {\bibfnamefont {S.}~\bibnamefont
  {Zhang}}, \bibinfo {author} {\bibfnamefont {Y.~G.}\ \bibnamefont {Ma}},
  \bibinfo {author} {\bibfnamefont {J.~H.}\ \bibnamefont {Chen}},\ and\
  \bibinfo {author} {\bibfnamefont {C.}~\bibnamefont {Zhong}},\ }\bibfield
  {title} {\bibinfo {title} {{Beam energy dependence of Hanbury-Brown-Twiss
  radii from a blast-wave model}},\ }\href
  {https://doi.org/10.1155/2016/9414239} {\bibfield  {journal} {\bibinfo
  {journal} {Adv. High Energy Phys.}\ }\textbf {\bibinfo {volume} {2016}},\
  \bibinfo {pages} {9414239} (\bibinfo {year} {2016})},\ \Eprint
  {https://arxiv.org/abs/1602.01564} {arXiv:1602.01564 [nucl-th]} \BibitemShut
  {NoStop}%
\bibitem [{\citenamefont {Landau}\ and\ \citenamefont
  {Lifshitz}(1980)}]{Landau:1980mil}%
  \BibitemOpen
  \bibfield  {author} {\bibinfo {author} {\bibfnamefont {L.}~\bibnamefont
  {Landau}}\ and\ \bibinfo {author} {\bibfnamefont {E.}~\bibnamefont
  {Lifshitz}},\ }\href@noop {} {\emph {\bibinfo {title} {{Statistical Physics,
  Part 1}}}},\ \bibinfo {series} {Course of Theoretical Physics}, Vol.~\bibinfo
  {volume} {5}\ (\bibinfo  {publisher} {Butterworth-Heinemann},\ \bibinfo
  {address} {Oxford},\ \bibinfo {year} {1980})\BibitemShut {NoStop}%
\bibitem [{Note5()}]{Note5}%
  \BibitemOpen
  \bibinfo {note} {Private communication with S.\ Floerchinger (2016), and
  unpublished work presented by S.\ Floerchinger at the workshop {"\protect
  \emph {Functional Methods in Strongly Correlated Systems}"} in Hirschegg,
  Austria, on April 4, 2019}\BibitemShut {NoStop}%
\bibitem [{\citenamefont {Braun-Munzinger}\ \emph {et~al.}(2003)\citenamefont
  {Braun-Munzinger}, \citenamefont {Redlich},\ and\ \citenamefont
  {Stachel}}]{BraunMunzinger:2003zd}%
  \BibitemOpen
  \bibfield  {author} {\bibinfo {author} {\bibfnamefont {P.}~\bibnamefont
  {Braun-Munzinger}}, \bibinfo {author} {\bibfnamefont {K.}~\bibnamefont
  {Redlich}},\ and\ \bibinfo {author} {\bibfnamefont {J.}~\bibnamefont
  {Stachel}},\ }\bibfield  {title} {\bibinfo {title} {{Particle production in
  heavy ion collisions}},\ }\href {https://doi.org/10.1142/9789812795533_0008}
  {\bibfield  {journal} {\bibinfo  {journal} {{Quark--Gluon Plasma 3}}\ ,\
  \bibinfo {pages} {491}} (\bibinfo {year} {2003})},\ \Eprint
  {https://arxiv.org/abs/nucl-th/0304013} {arXiv:nucl-th/0304013} \BibitemShut
  {NoStop}%
\bibitem [{\citenamefont {Zyla}\ \emph {et~al.}(2020)\citenamefont {Zyla} \emph
  {et~al.}}]{Zyla:2020zbs}%
  \BibitemOpen
  \bibfield  {author} {\bibinfo {author} {\bibfnamefont {P.~A.}\ \bibnamefont
  {Zyla}} \emph {et~al.} (\bibinfo {collaboration} {Particle Data Group}),\
  }\bibfield  {title} {\bibinfo {title} {{Review of Particle Physics}},\ }\href
  {https://doi.org/10.1093/ptep/ptaa104} {\bibfield  {journal} {\bibinfo
  {journal} {PTEP}\ }\textbf {\bibinfo {volume} {2020}},\ \bibinfo {pages}
  {083C01} (\bibinfo {year} {2020})}\BibitemShut {NoStop}%
\end{thebibliography}%

\end{document}